\ifx\pdfoutput\undefined
	\documentclass[aps,preprint,a4paper,twoside,onecolumn,superscriptaddress,
	showkeys,nofootinbib,amssymb,amsmath,amsfonts,showpacs]{revtex4}
\else
	\documentclass[aps,preprint,a4paper,twoside,onecolumn,superscriptaddress,
	showkeys,nofootinbib,amssymb,amsmath,amsfonts,showpacs,pdftex]{revtex4}
\fi

\usepackage{natbib}
\usepackage{amsmath}
\usepackage{feynmf}
\usepackage{amssymb}
\usepackage{graphicx}

\newcommand{\beq}{\begin{equation}}
\newcommand{\eeq}{\end{equation}}

\newcommand{\pder}[2]{\frac{\partial #1}{\partial #2}}

\newcommand{\Thc}{T_\mathrm{hc}}

\DeclareMathOperator{\eV}{eV}

\DeclareMathOperator{\MeV}{MeV}
\DeclareMathOperator{\GeV}{GeV}
\DeclareMathOperator{\Mpc}{Mpc}
\DeclareMathOperator{\Hz}{Hz}

\begin{document}

\title{On the interaction between thermalized neutrinos and cosmological gravitational waves above the electroweak unification scale}
\author{Massimiliano Lattanzi}
\affiliation{ICRA --- International Center for Relativistic Astrophysics.}
\affiliation{Dipartimento di Fisica, Universit\`a di Roma ``La Sapienza'', Piazzale Aldo Moro 2, I-00185 Roma, Italy.}
\author{Giovanni Montani}
\affiliation{ICRA --- International Center for Relativistic Astrophysics.}
\affiliation{Dipartimento di Fisica, Universit\`a di Roma ``La Sapienza'', Piazzale Aldo Moro 2, I-00185 Roma, Italy.}

\begin{abstract}We investigate the interaction between the cosmological relic neutrinos, and primordial gravitational waves entering the horizon before the electroweak phase transition, corresponding to observable frequencies today $\nu_0\gtrsim 10^{-5}\,\Hz$. We give an analytic formula for the traceless transverse part of the anisotropic stress tensor, due to weakly interacting neutrinos, and derive an integro-differential equation describing the propagation of cosmological gravitational waves at these conditions. We find that this leads to a decrease of the wave intensity in the frequency region accessible to the LISA space interferometer, that is at the present the most promising way to obtain a direct detection of  a cosmological gravitational wave. The absorbed intensity does not depend neither on the perturbation wavelength, nor on the details of neutrino interactions, and is affected only by the neutrino fraction $f_\nu$. The transmitted intensity amounts to $88\%$ for the standard value $f_\nu=0.40523$. An approximate formula for non-standard values of $f_\nu$ is given.
\end{abstract}

\maketitle

\section{Introduction}
The presence in the Universe today of a stochastic background of gravitational waves (GWs) is a quite general prediction of several early cosmology scenarios. In fact, the production of gravitons is the outcome of many processes that could have occurred in the early phases of the cosmological evolution. Notable examples of this kind of processes include the amplification of vacuum fluctuations in inflationary\cite{Gr75} and pre-big-bang cosmology scenarios\cite{Ga93}, phase transitions\cite{Ho86}, and finally the oscillation of cosmic strings loops\cite{Vi81}. In most of these cases, the predicted spectrum of gravitational waves extends over a very large range of frequencies; for example, inflationary expansion produces a flat spectrum that spans more than 20 orders of magnitude in frequency, going from $10^{-18}$ to $10^{9}$ Hz.

The detection of such primordial gravitational waves, produced in the early Universe, would be a major breakthrough in cosmology and high energy physics. This is because gravitons decouple from the cosmological plasma at very early times, when the temperature of the Universe is of the order of the Planck energy. In this way, relic gravitational waves provide us a ``snapshot'' of the Universe near the Planck time, in a similar way as the cosmic microwave background radiation (CMBR) images the Universe at the time of recombination.

The extremely low frequency (ELF) region ($\nu_0\lesssim10^{-15}\Hz$) in the spectrum of primordial gravitational waves can be probed through the anisotropies of the CMBR. In particular, gravitational waves leave a distinct imprint in the so-called magnetic or B-modes of its polarization field\cite{Hu97,Pr05}. The amplitude of the primordial spectrum of gravitational waves is usually parameterized through the tensor-to-scalar ratio $r$, i.e., the ratio between the amplitudes of the initial spectra of the tensor and scalar perturbations in the metric. The Planck satellite\cite{Planck}, scheduled for launch in 2007, is expected to be sensitive\cite{Tu05} to $r\ge0.05$. The lower limit corresponds to a density parameter $\Omega_{GW}(\nu)\equiv (1/\rho_c) d\rho_{GW}/d\log\nu $ as faint as $\sim 3\times10^{-16}h^{-2}$ ($h$ is the dimensionless Hubble constant) in the ELF range. Although this value looks incredibly small, it should be noted that, in order to produce such an amount in the framework of inflationary models (that at present time represent the most promising way to produce a signal in the region under consideration), a very early (starting at $t\sim10^{-38}$ sec) inflation is required, and this possibility looks, from a theoretical point of view, quite unnatural. Future experiments are expected to enhance the sensitivity of two orders of magnitude\cite{Tu05}.

On the other hand the planned large scale interferometric GW detectors, although designed with the aim to detect astrophysical signals, can possibly also detect signals of cosmological origin\cite{Ma00}. They give complementary information with respect of the CMBR polarization field since, even if their sensitivity is by no means comparable to the one than can be reached by CMBR polarization experiments, nevertheless they probe a different region in the frequency domain that would not be accessible to those ones. In particular the ground-based interferometers, such as the LIGO\cite{LIGO}, VIRGO\cite{VIRGO}, GEO600\cite{GEO600} and TAMA300\cite{TAMA300} experiments, operate in the range $1\,\Hz<\nu_0<10^4\,\Hz$, and are expected to be sensitive to $\Omega_{GW}h^2\ge 10^{-2}$. Even more interesting is the LISA space interferometer\cite{LISA}, that will probably operate from 2013 to 2018. Not being hampered by the Earth seismic noise, it will probe the frequency region between $10^{-4}$ and $1\,\Hz$ and will in principle be able to detect $\Omega_{GW}h^2\ge 10^{-12}$ at $\nu_0=10^{-3}\,\Hz$. According to theoretical predictions, a large enough GW signal at this frequencies can be produced, with the appropriate choice of parameters, by a pre-big-bang accelerated expansion, by the oscillation of cosmic strings, or by the electroweak phase transition occurring at $T=300\,\GeV$.

In order to compare the theoretical predictions with the expected instrument sensitivities, one needs to evolve the GWs from the time of their production to the present. It is usually assumed that gravitons propagate in vacuum, i.e., they freely stream across the Universe. In this case, the only effect on a propagating GW is a change in frequency (corresponding to the usual redshift of the graviton energy caused by the expansion of the Universe), while the intensity of the wave remains the same. However, GWs are sourced by the anisotropic stress part of the energy-momentum tensor of matter, so that the vacuum approximation is well-motivated only when this can be neglected. It is already known that the anisotropic stress of free streaming neutrinos acts as an effective viscosity, absorbing gravitational waves in the ELF region, thus resulting in a damping of the B-modes of CMBR\cite{Bo96,Du98,We04,Ba05}. The aim of this letter is to understand if something similar happens in other regions of the frequency domain. In particular, we consider GWs that enter the horizon before the electroweak phase transition (EWPT). This corresponds to an observable frequency today $\nu_0\gtrsim 10^{-5}\,\Hz$, i.e., to all waves possibly detectable by interferometers. In order to study this issue, we solve the Boltzmann equation for the distribution function of cosmological neutrinos and use its formal solution to obtain an integral expression for the anisotropic stress of neutrinos. We follow the approach outlined in Ref. \cite{We04}, but we have to take into account the interactions of neutrinos with other background particles. The anisotropic stress is then inserted as an integral source term in the Einstein equation for the evolution of tensor metric perturbations. In the limit of very short neutrino mean free path, relevant for waves well below a frequency of $10^8\,\Hz$, this equation can be cast in purely differential form and a numerical solution be sought with standard methods. We find that the intensity of GWs is reduced to $\sim 90\%$ of its value in vacuum, its exact value depending only on one physical parameter, namely the density fraction of neutrinos. Neither the wave frequency nor the detail of neutrino interaction affect the value of the absorbed intensity, resulting in an universal behaviour in the frequency range considered. 

The paper is organized as follows. In section \ref{sec:prop} basic statements are provided in order to formulate the dynamical problem; in particular the coupled Einstein-Boltzmann system is discussed in correspondence to the interaction between GW and neutrino plasma. Moreover, a detailed analysis of the perturbative anisotropic stress tensor together with the optical depth is presented. In section \ref{sec:abs}, we develop the real calculation of the GW absorption by neutrinos, by virtue of an approximation scheme of the source term, based on the fast interaction rate of neutrinos. The time behaviour of the transmitted GW intensity is numerically determined, and its dependence on the neutrino content of the Universe outlined. In section \ref{sec:conc}, concluding remarks follow in order to stress the phenomenological relevance of our analysis, with respect to GW interferometric detectors.

\section{Propagation of cosmological gravitational waves through a neutrino plasma}
\label{sec:prop}

\subsection{Basic formalism}

We shall use, all throughout the paper, natural units in which $c=\hbar=k_B=1$.

Let us consider a gravitational wave, propagating on the background of a flat Friedmann Universe. In synchronous gauge, the spatial components of the perturbed metric are written as:
\begin{equation}
g_{ij}=a^2(t)\left[\delta_{ij}+h_{ij}\right]
\end{equation}
while the other components are left unperturbed: $g_{00}=1$ and $g_{0i}=0$. Here $a(t)$ is the cosmological scale factor, and $h_{ij}$ is a small tensor perturbation representing a GW, that evolves according to\cite{We72}:
\begin{equation}
\partial^2_t{h}_{ij}+\left(\frac{3}{a}\frac{da}{dt}\right)\partial_t{h}_{ij}-\left
(\frac{\nabla^2}{a^2}\right)h_{ij}=16\pi G\pi_{ij}\;,
\label{eq:tens evol}
\end{equation}
where $\pi_{ij}$ is the anisotropic stress part of the energy-momentum tensor $T^{\mu}_{\phantom\mu\nu}$. It is defined through the relation $\pi_{ij}=T^i_{\phantom ij}-{\cal P}\delta_{ij}$, where $\cal P$ is the background pressure.
The space part of the energy momentum tensor can be computed by means of:
\begin{equation}
T^i_{\phantom ij}=\frac{1}{\sqrt{-g}}\int f(x^i, p_j,t) \frac{ p^i p_j}{p^0}\,dp_1\,dp_2\,dp_3.
\label{eq:en-mom tens}
\end{equation}
where $g$ is the determinant of the metric, $p^\mu\equiv dx^\mu/d\lambda$ is the four-momentum ($\lambda$ is an affine parameter), and $f(x^i, p_j,t)$ is the distribution function (DF) of the particles in the fluid. The DF evolves according to the Boltzmann equation:
\begin{equation}
\hat L[f]=\hat C[f]
\label{eq:boltz}
\end{equation}
where the $\hat L\equiv df/d\lambda$ is the Liouville operator, and $\hat C$ is the collision operator accounting for collisions between particles. Using the geodesic equation, the left-hand side (LHS) of the Boltzmann equation (the so called Liouville operator $\hat L$) for $f$ can be cast in the form (summation over repeated indices is understood):
\begin{equation}
\hat L[f]\equiv\frac{df}{d\lambda}=p^0\bigg[\pder{ f}{t}+\frac{p_i}{p^0}\pder{f}{x_i} 
+\frac{p^j p^k}{2 p^0}\pder{g_{jk}}{x_i}\pder{f}{p_i}\bigg].
\end{equation} 
Once the collision term is also specified, eqs. (\ref{eq:tens evol}), (\ref{eq:en-mom tens}) and (\ref{eq:boltz}), together with the Friedmann equation describing the background evolution, are all that is needed, at least in principle, to follow the propagation of a GW.

\subsection{Anisotropic Stress Computation}

In this letter we shall consider GWs that enter the Hubble horizon well before neutrino decoupling (occurring at $T\sim 1\MeV$), and we shall suppose that the main contribution to the anisotropic stress of matter is given by neutrinos \footnote{This approximation relies in the fact that, roughly speaking, the anisotropic stress is proportional to the mean free path of the particles in the fluid. In the framework of the standard model of particle physics, neutrinos are the most weakly interacting particles in the Universe, and then their anisotropic stress is larger with respect to other particle species.}. We can write the DF of neutrinos as:
\begin{equation}
f(x^i, p_j,t)=f(x^i, p_j, t_1) + \delta f(x^i, p_j,t),
\end{equation}
so that at the initial time $t_1$ we have $\delta f=0$. We take for the DF at the initial time the usual Fermi-Dirac form
\footnote{Please note that this is not the usual splitting into a zeroth-order, equilibrium part plus a first-order part, since $f(t_1)$ contains itself a first-order part. However, $\delta f$ is certainly first order since the zeroth-order part of the DF (\ref{eq: FermiDirac}) satisfies the collisionless Boltzmann equation over the flat, unperturbed Robertson-Walker metric.}:
\begin{equation}
f(x^i, p_j,t_1)=\frac{N}{(2\pi)^3}\left[\exp\left(\sqrt{g_{ij}p_i p_j}/T\right)+1\right]_{t=t_1},
\label{eq: FermiDirac}
\end{equation}
where $N$ counts the number of internal degrees of freedom.

In respect of the case of GWs entering the horizon \emph{after} neutrino decoupling (as in Ref. \cite{We04}), the situation we are going to deal with is complicated by the fact that the right-hand side (RHS) of the Boltzmann equation does not vanish, neutrinos being yet interacting with the cosmological plasma. The calculation of the exact collision term is usually not easy to deal with; however, a reasonable guess is given by assuming that $\hat C[\delta f]=-p^0\delta f/t_c$, where $t_c$ is the mean time between collisions, and the $p^0$ factor comes from writing the Boltzmann equation in terms of the conformal parameter $\lambda$.

The following relations will be useful in the following (we define $p\equiv\sqrt{p_i p_i}$):
\begin{equation}
p^0=\frac{p}{a}+O\,(h_{ij}),\quad p^i=\frac{p_i}{a^2}+O\,(h_{ij}),
\end{equation}
We Fourier transform the spatial dependence of the involved quantities. The comoving wavenumber $k_i$ is the Fourier conjugate variable to the comoving coordinate $x^i$. Introducing the time variable $u$:
\begin{equation}
u \equiv k \int_{t_1}^t\frac{dt'}{a(t')}, 
\end{equation}
the Boltzmann equation finally reads, to first order in perturbed quantities:
\begin{equation}
\dot \delta f-i\,\hat p_i \hat k_i\delta f -\frac{i\,p}{2} \hat p_i\hat p_j 
\hat p_k \hat k_i\,\pder{f^{(0)}}{p} \left[ h_{jk}(u)- h_{jk}(0) \right] 
=-\frac{a}{k}\cdot\frac{\delta f}{t_c}
\end{equation}
where the dot denotes derivative with respect to $u$, and hats denote unit vectors; $f^{(0)}$ is the zeroth-order part of $f$. Defining:
\begin{subequations}
\begin{align}
&J(u,k_i,p_j)\equiv - i \hat p_i \hat k_i + \frac{a}{k t_c},\\
&\alpha_{ij}(u,k_i,p_j)\equiv \frac{ip}{2} \hat k_l \hat p_l \hat p_i \hat p_j\,\pder{f^{(0)}}{p},
\end{align}
\end{subequations}
a formal solution for $\delta f$ satisfying the initial condition $\delta f(0)=0$ can be written as (for shortness of notation we keep indicating only the dependence on $u$):
\begin{align}
\delta f(u) = \exp\left[-\int_0^u J(u')du' \right]\times
\int_0^u \exp \left[\int_0^{u'} J(u'')du''\right]
\alpha_{ij}(u')\Big[ h_{ij}(u') - h_{ij}(0) \Big]du'.
\raisetag{2pt}
\end{align}
Now the anisotropic stress tensor can be computed by means of equation (\ref{eq:en-mom tens}); the result is:
\begin{align}
\pi_{ij}=-4 \rho_\nu\int_0^u\bigg\{\dot h_{ij}(u')+\dot \tau(u')\Big[h_{ij}(u')-h_{ij}(0)\Big]\bigg\}\times
\,
e^{\tau(u')-\tau(u)}K(u-u')\,du,
\label{eq:an stress}
\end{align}
where $\rho_\nu$ is the mean (background) neutrino energy density, $K$ is the kernel:
\begin{equation}
K(s)\equiv\frac{1}{16}\int_{-1}^{+1}(1-x^2)e^{ixs}\,dx,
\end{equation}
and $\tau$ is the optical depth of the plasma with respect to the propagation of neutrinos:
\begin{equation}
\tau(u)=\int_{t_1}^{t}\ n \sigma t\,dt=\frac{1}{k}\int_0^u a  n\sigma\,  du.
\end{equation}
Here $\sigma$ is the relevant cross section for the reactions that keep neutrinos coupled to the other particles in the plasma, $n$ is the density of target particles. We note that since $n \sigma= t_c$, we have $\dot \tau = a/ kt_c$. The dimensionless quantity $\tau$ gives an estimate of the average number of interactions occurring between $t_1$ and $t$.

\subsection{Behaviour of the optical depth}
In order to proceed further, we need to calculate the optical depth $\tau(u)$. In the following we shall consided GWs entering the horizon before the EWPT ($T\sim~300\,\GeV$). In order to better understand which region in frequency space we are investigating when we make this assumption, it is useful to find the relation between the gravitational wave frequency and the temperature of the Universe at the time when its horizon crossing occurs. This can be done very easily as follows. The temperature $\Thc$ at which a wave, having today a physical wavelength equal to $\lambda_0$, entered the horizon is given by \cite{Ko90}:
\begin{equation}
\Thc(\lambda_0)=63 g_*^{-1/2}g_{*S}^{1/3}\left(\frac{\lambda_0}{\Mpc}\right)^{-1}\,\eV,
\label{eq:Thc}
\end{equation}
where $g_*$ and $g_{*S}$ measure the contribution of relativistic species to the energy density and entropy density respectively. Before EWPT, all species in the standard models of fundamental interactions are relativistic, resulting in $g_*=106.75$. Furthermore, in thermal equilibrium $g_{*S}=g_*$. Then we find that the present day frequency $\nu_0$ of the wave is:
\begin{equation}
\nu_0=3.36\times 10^{-5}\left(\frac{\Thc}{100\,\GeV}\right)\,\Hz
\end{equation}
Quite interestingly, we find that all cosmological GWs that can be detected by laser interferometers, both spatial and ground-based, entered the horizon prior to the electroweak symmetry breaking.

Neutrinos are kept in thermal equilibrium with the cosmological plasma through the weak interactions with other particles. At energies large as the ones we are considering, all the particles in the standard model of interactions are a possible target for neutrinos. The density $n$ that appears in the expression for $\tau$ is then
\begin{equation}
n=\frac{\zeta(3)}{\pi^2}g_{*N} T^3
\label{eq:n lepton}
\end{equation}
where $\zeta(3)$ is the Riemann zeta function of order 3, and $g_{*N}$, in a similar way than $g_*$ and $g_{*S}$, measures the contributions of relativistic species to the particle number density. At $T\gtrsim300\,\GeV$, $g_{*N}=95.5$. This yields $n\simeq11.6\,T^3$.

For what concerns the cross section, we have that in the regime corresponding to average thermal energies much larger than the rest masses of the $W^\pm, Z^0$ weak bosons $m_w,\,m_z\sim 100\,\GeV$ the temperature dependence of $\sigma$ is roughly given by:
\begin{equation}
\sigma \sim \displaystyle{\frac{G_F^2 m_w^4}{T^2}} \qquad (T\gg m_w)
\label{eq:cross section}
\end{equation}
where $G_F=1.1664\times10^{-5}\,\GeV^{-2}$ is the Fermi coupling constant.
More detailed calculations can be performed to bring more precise formulas for the cross section. However, as we shall see in the following, the above expression us enough for our purposes.

The last step in the computation of $\tau$ is to express the comoving wavenumber $k$ in terms of $\Thc$. Considering the initial time $t_1$ to be very early, so that $t_1\simeq 0$, we have that in the radiation dominated era $a(t)\propto\sqrt{t}$ and $a(u)\propto u$. Moreover, since $\int_0^t dt'/a(t')$ is the comoving horizon size, we have that $u=1$ corresponds to the time of horizon crossing. We choose to normalize the scale factor so that it is equal to unity at horizon crossing, then we take $a(u)=u$. Now we just need to consider the physical wavelength $\lambda$ as a function of temperature\cite{Ko90}:
\begin{equation}
\lambda(T)=5.83 \times 10^{25} \left(\frac{\lambda_0}{\Mpc}\right)
\cdot\left(\frac{T}{\GeV}\right)^{-1} g_{*S}^{-1/3} \GeV^{-1}.
\end{equation}
Evaluating this expression at $\Thc$ and eliminating $\lambda_0$ by means of eq. (\ref{eq:Thc}), we get for the comoving wavenumber $k$:
\begin{equation}
k=2.76\cdot 10^{21} \left(\frac{\Thc}{\GeV}\right)^2\Mpc^{-1}.
\end{equation}
Using this expression and eqs. (\ref{eq:n lepton}) and (\ref{eq:cross section}), together with fact that the quantity $aT g_*^{1/3}$ is constant we find, for $T\gg 100\,\GeV$:
\begin{equation}
\tau(u)\simeq 3\times 10^{13}\left(\frac{\Thc}{100 \mathrm{GeV}}\right)^{-1}u.
\label{eq:tau}
\end{equation}

\section{Absorption of cosmological gravitational waves}
\label{sec:abs}

Now we can finally deal with the GW evolution equation. Defining $\chi(u)\equiv h_{ij}(u)/h_{ij}(0)$,
the equation reads:
\begin{equation}
\ddot \chi+2{\cal H} \dot \chi + \chi=-24 f_\nu	{\cal H}^2 \times 
\int_0^uK(u-u')\bigg[\dot \chi'+\dot \tau'\Big(\chi'-1\Big)\bigg]
e^{\tau'-\tau}du.
\label{eq:chi intdiff}
\end{equation}
Here $\cal H$ is the ``Hubble constant'' $\dot a/a= u^{-1}$, $f_\nu$ is the ratio of neutrino density to the total density (in the standard model, $f_\nu = 0.40523$), and $\tau$ is given by eq. (\ref{eq:tau}). We use the short-hand notation that primed quantities are evaluated in $u'$, and unprimed ones are evaluated in $u$. The above equation is an integro-differential equation for $\chi$. It is supplied by the initial conditions \cite{We04}:
\begin{equation}
\begin{array}{l}
\chi(0)=1, \\
\dot \chi(0)=0.
\end{array}
\end{equation}
A numerical solution can be found by using standards method for integro-differential equations\cite{De85,Li85}. It is however useful to consider two limiting cases. For waves entering the horizon very early (i.e, $\Thc\gg 10^{15} \GeV$, or equivalently $\nu_0\gg 10^8 \Hz$), the derivative of the optical depth is very small and then Eq. (\ref{eq:chi intdiff}) recover the free streaming form found in Ref. \cite{We04}. It is found in that paper that this results in an absorption of 35\% of the wave squared amplitude. On the opposite, for waves having $\Thc\ll 10^{15}\,\GeV$, the derivative of the optical depth is very large. This makes the integration of the evolution equation not trivial, since the integral on the right hand side receives contribution only from a very small region where $e^{\tau'-\tau}$ significantly differs from zero. Technically speaking, the integral kernel is said be ``quasi singular''\cite{De85,Li85}.  However, we can exploit this characteristic to our advantage. In fact, when the following disequality holds:
\begin{equation}
\left|\frac{\dot \chi}{\chi-1}\right|\ll|\dot\tau|
\label{eq:tight condition}
\end{equation}
we can approximate $e^{\tau-\tau'}$ with a Dirac delta function $\delta(\tau-\tau')$. The above condition can be checked \emph{a posteriori} after obtaining a solution. However, apart from the mathematical details, it is clear that this approximation relies on the fact that the characteristic time scale for the variation of the gravitational field is much larger than the mean time between collisions. The integral source term on the RHS of eq. (\ref{eq:chi intdiff}) depends, in general, on the whole past history of the system. Nevertheless, when the condition expressed in (\ref{eq:tight condition}) holds, the very fast interaction rate destroys any information on the past history of the system, so that the integral is dominated by local (in time) contributions.

This condition can, on the other hand, tested \emph{a priori} using a test function in place of the actual solution $\chi(u)$. A reasonable guess is to suppose that the solution we are seeking does not differ too much from the vacuum solution $\chi_{\mathrm{vac}}$, i.e. the solution in absence of anisotropic stress. The solution of eq.(\ref{eq:chi intdiff}) with vanishing RHS is $\chi_{\mathrm vac}(u)=\sin(u)/u$. We use this as a test function. Inside the horizon ($u\gtrsim 1$), where the source term is really effective\footnote{Well outside the horizon, $\dot \chi_\mathrm{vac}\simeq 0$ and $\chi_\mathrm{vac}\simeq 1$, so that the source term is very small.}, we have that $\big|\dot \chi_{\mathrm{vac}}/({\chi_{\mathrm{vac}}-1)}\big|<1$. The condition \eqref{eq:tight condition} then simply reads $\dot\tau\gg 1$. 

Seen in another way, the expression of the vacuum solution in terms of its Fourier components is:
\begin{equation}
\chi_{\mathrm vac}(u)=\frac{1}{2}\int_{-1}^{1}e^{-i u v}dv 
\end{equation}
so that only harmonics with ``frequency'' $|v|\le 1$ give a contribution to this solution. The condition that the mean collision time is much smaller than the time scale for the GW evolution, reads then $\dot\tau\gg v\ge1$ . Comparing with eq. (\ref{eq:tau}), we see that this is the case for all $k$ modes entering the horizon before the EWPT, up to modes entering as earlier as $\Thc\sim 10^{15} \GeV$.

Thus, using the substitution:
\begin{equation}
e^{\tau'-\tau}\to\delta(\tau'-\tau)=\frac{\delta(u'-u)}{|\dot \tau|},
\end{equation}
 the evolution equation reads:
\begin{equation}
\ddot \chi+\frac{2}{u} \dot \chi + \chi=-\frac{8 f_\nu}{5 u^2} (\chi-1)
\label{eq:chi diff}
\end{equation}
where we have dropped a term proportional to $\dot \chi/\dot\tau$.

This ordinary differential equation, describing the evolution of a gravitational wave in a tightly coupled plasma, is, together with the more general equation (\ref{eq:chi intdiff}) and the expression (\ref{eq:an stress}) for the anisotropic stress, the main achievement of this letter. We note that, once the condition (\ref{eq:tight condition}) is fulfilled, the GW evolution does not depend on the details of the interaction, since the optical depth, and consequently the interaction cross section, do not appear anymore. The only physical parameter related to the background conditions appearing in eq. (\ref{eq:chi diff}), is the fraction of energy density contributed by neutrinos. Moreover, using the $u$ time parameterization, the equation is universal in the sense that it does not depend on the wavenumber $k$.

\begin{center}
\begin{figure}
\resizebox{\textwidth}{!}{\includegraphics{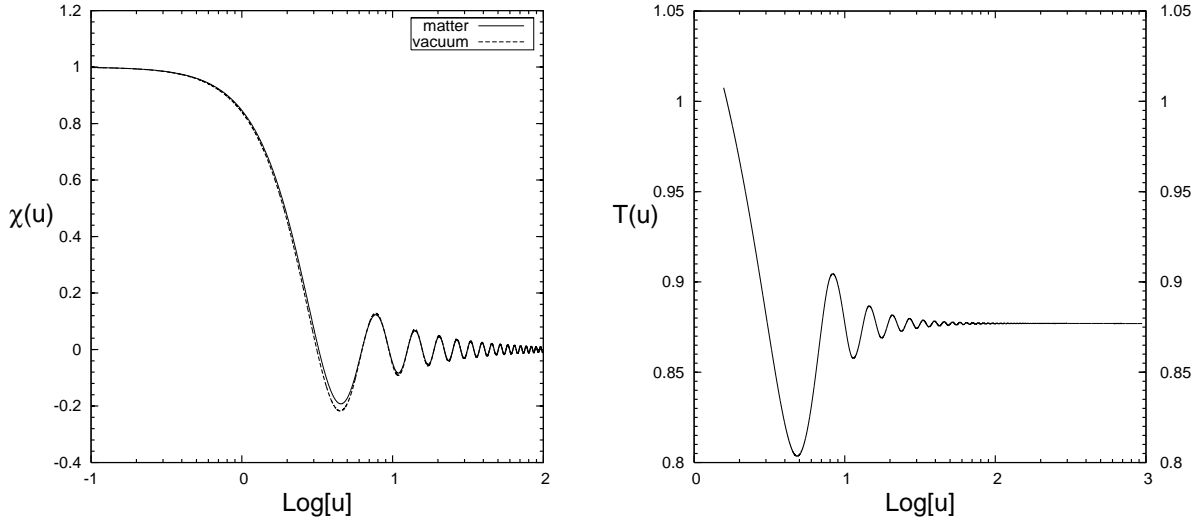}}
\caption{(\emph{Left panel}) Time evolution of the gravitational wave amplitude $\chi(u)$. Solid line represents a GW propagating in neutrino matter. Dashed line represents a GW propagating in vacuum. (\emph{Right panel}) Time evolution of the transmitted wave intensity ${\cal T}$.}
\label{fig: chi+T}
\end{figure}
\end{center}

A numerical solution to the above equation can be found quite easily. We plot this solution (let us it call the ``matter solution'') for the standard value $f_\nu=0.40523$, together with $\chi_{\mathrm vac}$, in the left panel of Fig. \ref{fig: chi+T}. The solution follows quite closely the vacuum one. A closer inspection shows that, far after the horizon crossing, the squared amplitude of a wave propagating in the cosmological plasma gets damped. We use as a measure of the transmitted intensity the following quantity ${\cal T}$:
\begin{equation}
{\cal T}\equiv \frac{\langle\chi^2\rangle}{\langle\chi_\mathrm{vac}^2\rangle},
\end{equation}
where brackets denote time average over a period.
In particular, we find that at $u=100$, ${\cal T}\simeq 0.9$, so that the intensity is reduced by a factor $\sim 10\%$ with respect to the vacuum case. We expect the damping not to become too much larger than this value, since the source term is multiplied by a factor $1/u^2$, and the gets rapidly suppressed after the entrance in the horizon. This is confirmed by examining the behaviour in time of ${\cal T}$, shown the right panel of Fig. \ref{fig: chi+T}. We see that it approaches asimptotically a constant value, let us denote it with ${\cal T}_{\infty}$, equal to $0.88$.

Non-standard physics (for example, the presence of a lepton asymmetry) can result in a different value for $f_\nu$ and consequently for the transmitted intensity ${\cal T}_\infty$. Integrating the equation for	different values of $f_\nu$ in the range $[0,1]$ we find that, with a very good approximation, ${\cal T}_\infty$ depends linearly on $f_\nu$. A fitting formula is given by:
\begin{equation}
{\cal T}_\infty=1-0.32 f_\nu + 0.05 f_\nu^2
\end{equation}

\section{Concluding Remarks}
\label{sec:conc}

Our analysis outlined how gravitational waves produced in the early Universe are softly damped due to the anisotropic stress of cosmological background neutrinos. We have considered gravitational waves entering the horizon before the electroweak phase transition, corresponding to observable frequencies today $\nu_0\gtrsim 10^{-4} \Hz$. We find that the anisotropic stress is closely related to the optical depth $\tau$ of the cosmological plasma with respect to the propagation of neutrinos, and to its first derivative with respect to time. In the limit of very small time derivative of the optical depth, relevant for frequencies $\nu_0\gg 10^{8}\,\Hz$, the behaviour studied in Ref. \cite{We04} is recovered, resulting in an absorption of 35.6\% of the wave intensity. This frequencies are however unaccessible to present-day interferometers. In the opposite limit of very large values of this derivative, relevant for observables frequencies $\nu_0\ll 10^{8}\,\Hz$, we find that roughly $12\%$ of the wave intensity is absorbed by the cosmological plasma, independently from its wavelength. This result does not depend on the details of the weak interactions between particles in the plasma, but depends only on the amount of neutrinos present in the Universe, as parameterised by the quantity $f_\nu$.  

The importance of our results relies in the fact that the damping affects GWs in the frequency range where the LISA space interferometer and future, second generation ground-based interferometers can possibly detect a signal of cosmological origin. This effect is roughly of the same order of magnitude as the one affecting GWs detectable through the B-modes of CMB polarization. The damping is not so severe to make the detection of cosmological waves unfeasible by interferometers. However it should be taken into account when testing the theoretical predictions of early Universe scenarios against observations. Moreover, the dependence of $T_{\infty}$ on $f_\nu$ can be exploited to measure the latter, and to constrain models of non-standard physics. This even more important in view of the fact that in this way we could measure the value of $f_\nu$ at very early times, while available constraints regard the neutrino fraction at the time of cosmological nucleosynthesis or at the time of matter-radiation decoupling.

\subsection*{Acknowledgments}

The authors would like to thank Remo Ruffini for having attracted our attention on this problem and for his valuable advice about its treatment.

\end{document}